\documentclass[12pt]{iopart}
\usepackage{color}
\usepackage{graphicx}
\usepackage{bm}
\usepackage{braket}

\begin{document}

\title{QFT based quantum arithmetic logic unit on IBM quantum computer}

\author{Sel\c{c}uk \c{C}akmak$^{1,*}$, Murat Kurt$^2$ and Azmi Gen\c{c}ten$^2$}

\address{$^1$ Department of Software Engineering, University of Samsun, 55420 Samsun, T\"{u}rkiye}
\address{$^2$ Department of Physics, Ondokuz May{\i}s University, 55139 Samsun, T\"{u}rkiye}
\address{$^{*}$ Author to whom any correspondence should be addressed.}
\ead{selcuk.cakmak@samsun.edu.tr}

\begin{abstract}
In this study, we constructed a primitive quantum arithmetic logic unit (qALU) based on the quantum Fourier transform. The qALU is capable of performing arithmetic ADD (addition) and logic NAND gate operations. We presented two versions of the arithmetic logic unit, with two-input and four-input capabilities. By comparing the required number of quantum gates for serial and parallel architectures in executing arithmetic addition, we evaluated the performance. We executed our quantum Fourier transform based qALU design on real quantum computer hardware provided by IBM. The results demonstrated that the proposed circuit can perform arithmetic and logic operations with a high success rate. Furthermore, we discussed the potential implementation of qALU circuit in the field of computer science, highlighting the possibility of constructing soft-core processor on quantum processing unit.
\end{abstract}

\section{Introduction}
Quantum computing, a revolutionary field at the intersection of physics and computer science, harnesses the principles of quantum mechanics and has the potential to overcome the limitations of classical computing systems, also known as quantum supremacy~\cite{Preskill18,Zhong20}.
In recent years, programmable noisy intermediate-scale quantum (NISQ) devices~\cite{Murali19,Murali20,Cheng23,Niu23}, based on superconducting~\cite{Kwon21,Krantz19,Gambetta17} and trapped ion~\cite{Kielpinski02,Moses23,Blumel21} qubit technologies, have become accessible through cloud-based quantum computing services such as IBM quantum composer~\cite{IBMqc}.

Quantum computer is capable of performing any computation that a conventional computer can execute as well. The arithmetic and logic (Boolean) operations are executed in arithmetic logic unit (ALU) in central processing unit (CPU) of the conventional computers~\cite{Pacheco21,Catsoulis05}. In addition to reputed quantum algorithms, such as Deutsch-Jozsa~\cite{Deutsch92}, the capability of quantum information allows quantum computing systems to perform both arithmetic and logic operations through the use of quantum logic gates. In recent studies, the quantum circuits have been designed to realize some arithmetic and logic operations~\cite{Thomsen10,Ayyoub17,Bolhassani17}. However, it is not easy to scale these circuits for large input bits (i.e., large numbers) or multiple inputs.

The quantum version of the discrete Fouirer transform called as quantum Fourier transform (QFT) is the critical process of period finding (quantum phase estimation) based algorithms, such as Shor's factorization~\cite{Shor94} and HHL algorithm~\cite{Harrow09}. On the other hand QFT enables to develop new approachs in the filed of quantum information processing. Specially, QFT provides downsizing the quantum circuit that implements arithmetic operations~\cite{Draper00}. Moreover, operating in the Fourier basis reduces the computational cost and accelerates the processing time. However, recent studies have presented arithmetic addition, subtraction, and multiplication operations using QFT-based quantum circuits~\cite{Draper00,Perez17,Sahin20,Pavlidis21}. Thus far, no studies have explored arithmetic logic units based on the quantum Fourier transform.

In this research, we propose and implement a quantum fourier transform based primitive quantum arithmetic logic unit which drives only an arithmetic addition (ADD) and a logic NOT-AND (NAND) gate operations. We demonstrate the proposed quantum arithmetic logic unit algorithm on real quantum computer. We also compare the required number of quantum logic gates to realize arithmetic operations for two-input and multi-input architectures on quantum processing unit. We will discuss the results and possible implementations of the proposed quantum circiut.

\section{Motivation}
Quantum computer is an universal computer in the sense that it can theoretically perform any computation that a classical computer can do. This is because quantum computers are not limited to the binary system of classical computers and can use quantum states (quantum register) to represent information, which allows for parallel computation and more fast and efficient algorithms.

The arithmetic operations (addition, subtraction, multiplication, and division) are combinational digital circuits consisting of elementary logic gates (such as AND, OR, XOR). In classical information processing, both arithmetic and logic operations are performed by the ALU. Moreover, the ALU is a key component of central processing units. So, we are researching the possible implementations of the QFT-based qALU in the field of computer science. We aim to construct a quantum circuit on a quantum processor that performs arithmetic and logic operations based on quantum logic gates in the quantum Fourier space (basis). Here, the main motivation is to directly execute any classical computer program instructions on a quantum processor. So, we focus the developing to quantum arithmetic logic unit which is the common and most critical part of the central processing units (CPUs). The possible implementations will be discussed in detail in Sec.~\ref{conc}

\section{Preliminaries}
The theory of the quantum information processing is well-defined in literature~\cite{Nielsen12,Yanofsky08}. In quantum computing systems, the information of the classical data (represents the numbers) is encoded on the qubit by quantum state preparation protocol~\cite{Ashhab22}. In the technique of amplitude-embedding, data (classical information) is encoded by manipulating the amplitudes of a quantum state~\cite{Schuld18,Havlicek19}. Then the information is processed by the appled quantum logic gates~\cite{Nielsen12} to realize desired algorithm. In this part, we will only mention the short theory of the considered scheme in this paper.
\subsection{Quantum Fourier transform}
\begin{figure}[!ht]\centering
\includegraphics[width=.6\linewidth]{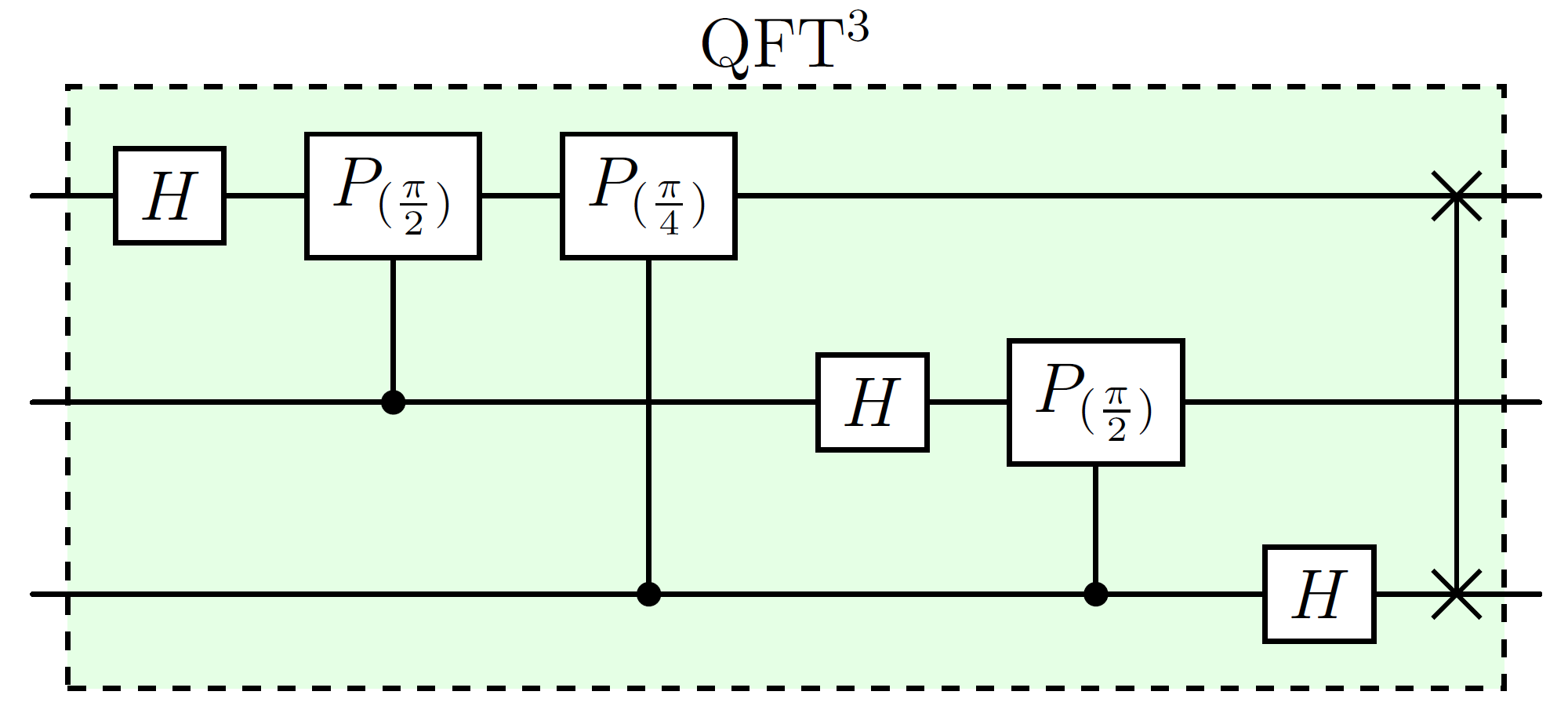}
\caption{\label{fig0} 3-qubit quantum Fourier transform circuit ($\mathrm{QFT^{n}}$ where $\mathrm{n=3}$ represents the number of qubits).}
\end{figure}

The quantum Fourier transform is the quantum analog of the discrete Fourier transform, which acts on the quantum state. Information processing is realized on the qubits, which are generated in a superposition within Fourier space (basis). This superposition enables the parallel processing of multiple computations, exponentially increasing the computational power of quantum systems. This is because the QFT requires $O(n^{2})$ operations for $n$ qubits system, while the classical FFT (Fast Fourier Transform) requires $O(n 2^{n})$ operations for $n$ bits. So, the opeations in FFT exponentially increases in comparison to the QFT. Basically, QFT is expressed by the following equation~\cite{Asaka00,Dixit22}
\begin{equation}\label{QFT}
\mathrm{QFT \ket{x}}=\frac{1}{\sqrt{N}}\sum_{y=0}^{N-1} e^{2\pi i xy/N}\ket{y}
\end{equation}
where the number of basis states is $N=2^n$, $\ket{x}=\ket{x_1} \otimes \ket{x_2} \otimes \ket{x_3} \cdots \otimes \ket{x_n}$ is in computational basis and $\ket{y}=\ket{y_1} \otimes \ket{y_2} \otimes \ket{y_3} \cdots \otimes \ket{y_n}$ is in Fourier basis. QFT can be applied by quantum circuit only includes the Hadamard ($\mathrm{H}$), conditional (or controlled) phase-shift ($\mathrm{P}$) and $\mathrm{SWAP}$ quantum logic gates. The generalized form of QFT algorithm can be found in Ref~\cite{Cao11}. The 3-qubit QFT circuit is presented in Fig.~\ref{fig0} that we are using it in Sec.~\ref{res}.

\subsection{Arithmetic logic unit}
Arithmetic logic unit (ALU) is a fundamental component of the central processing unit (conventional CPUs) and is responsible for performing basic arithmetic (ADD: addition, SUB: subtraction, MULT: multiplication, DIV: division, SRL/R: shift left/right) and logic (NOT, AND, OR, NAND, NOR, XOR, XNOR) operations. It is first described as a part of von-Neumann arcitecture in 1945~\cite{Pacheco21,Neumann93}. The ALU is a critical component of any computer processor, and its speed and efficiency can greatly impact the overall performance of a computer system. Conventional (classical) computer CPUs uses one of a few common instruction set architectures such as popular instruction set architectures MIPS, RISC, ARM, x86. Basically, the CPU decode the program instructions and executes the arithmetic and logic operations on ALU~\cite{Patterson14}.

On the other hand, the quantum arithmetic logic unit (qALU) performs the arithmetic and logic operations using the quantum logic gates. In particular, we construct quantum Fourier transform based qALU to perform operations in Fourier space. This provides parallel processing on the quantum states.
\section{\label{res}Results}
In this part, we have constructed the quantum circuit that realizes arithmetic ADD and logic NAND operations based on the quantum Fourier transform. Such a quantum circuit is the primitive quantum analog of the conventional arithmetic logic unit (ALU) that we refer to as the QFT-based qALU.
\subsection{QFT based one-bit 2-input qALU}
\begin{figure}[!ht]\centering
\includegraphics[width=.9\linewidth]{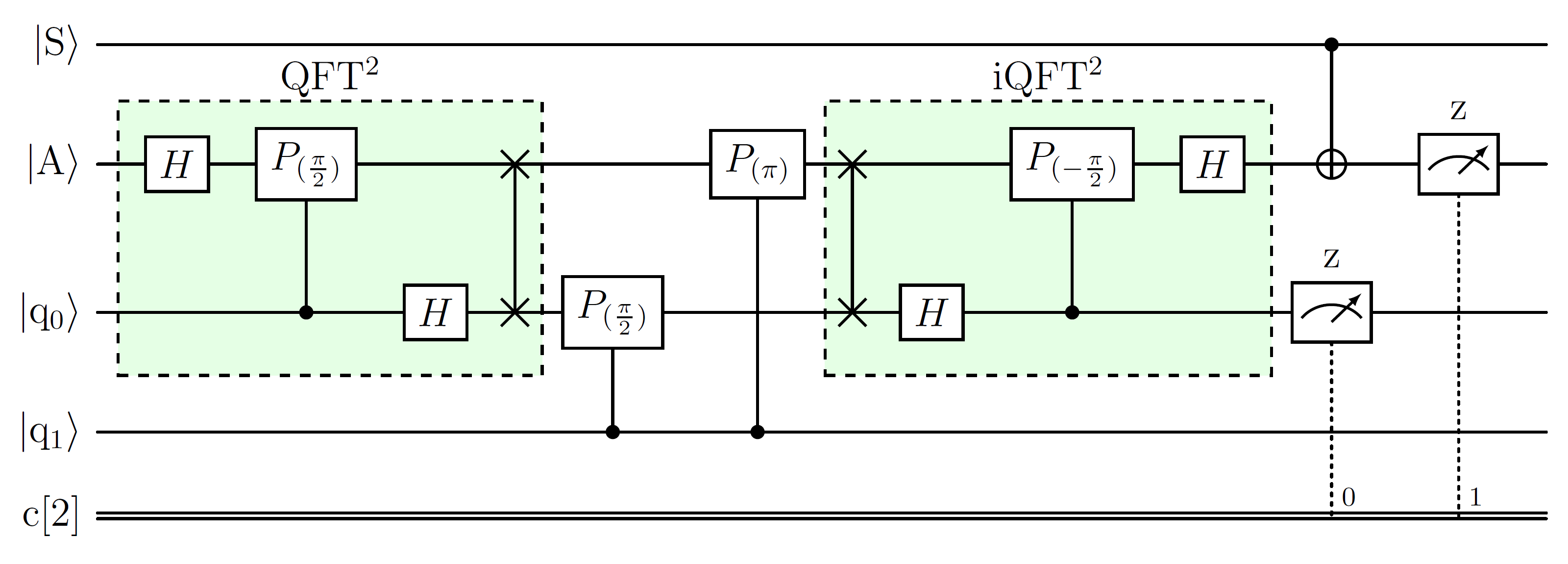}
\caption{\label{fig1} The quantum circuit of the quantum Fourier transform based 2-input qALU. Each input quantum channel holds one-bit information encoded in a qubit. The qALU performs arithmetic ADD (addition) or logic NAND gate operations when $\mathrm{\ket{S}=\ket{0}}$ or $\mathrm{\ket{S}=\ket{1}}$, respectively.}
\end{figure}

In Fig.~\ref{fig1}, we construct the quantum Fourier transform based qALU that performs an arithmetic ADD (addition) and a logic NAND (NOT-AND) gate on two qubits, each qubit holds the encoded data of classical bit. The quantum circuit includes four qubits. The qubit $\mathrm{\ket{S}}$ serves as a control for operation selection in the quantum arithmetic logic unit. When $\mathrm{\ket{S}=\ket{0}}$, the qALU executes the arithmetic ADD operation, while $\mathrm{\ket{S}=\ket{1}}$ signifies the execution of the logic NAND gate by the QFT-based qALU. $\mathrm{\ket{q_0}}$ and $\mathrm{\ket{q_1}}$ qubits are two quantum channels, each carrying the encoded information of a classical bit to be processed. Additionally, the qALU circuit requires a free qubit named ancillary, denoted as $\mathrm{\ket{A}}$ (which is set $\mathrm{\ket{0}}$ as default), to perform the algorithm. The qALU algorithm is performed in five steps as following: \textit{Step 1.} the quantum Fourier transform is applied on the $\mathrm{\ket{q_0}}$ and ancillary qubits. \textit{Step 2.} the desired operation is satified via controlled phase-shift gates. \textit{Step 4.} the inverse QFT is applied on the $\mathrm{\ket{q_0}}$ and ancillary qubits. \textit{Step 5.} the CNOT gate is applied to line up the results for ADD or NAND operations. Then, measurement process realized on the $\mathrm{\ket{q_0}}$ and $\mathrm{\ket{A}}$ qubits. Finally, the classical registers, $\mathrm{c[2]}$, will appear as $c_0$ holds the sum and $c_1$ holds the carry for arithmetic ADD operation or $c_1$ holds the result of NAND gate.

\begin{figure}[!ht]\centering
\includegraphics[width=6.5cm]{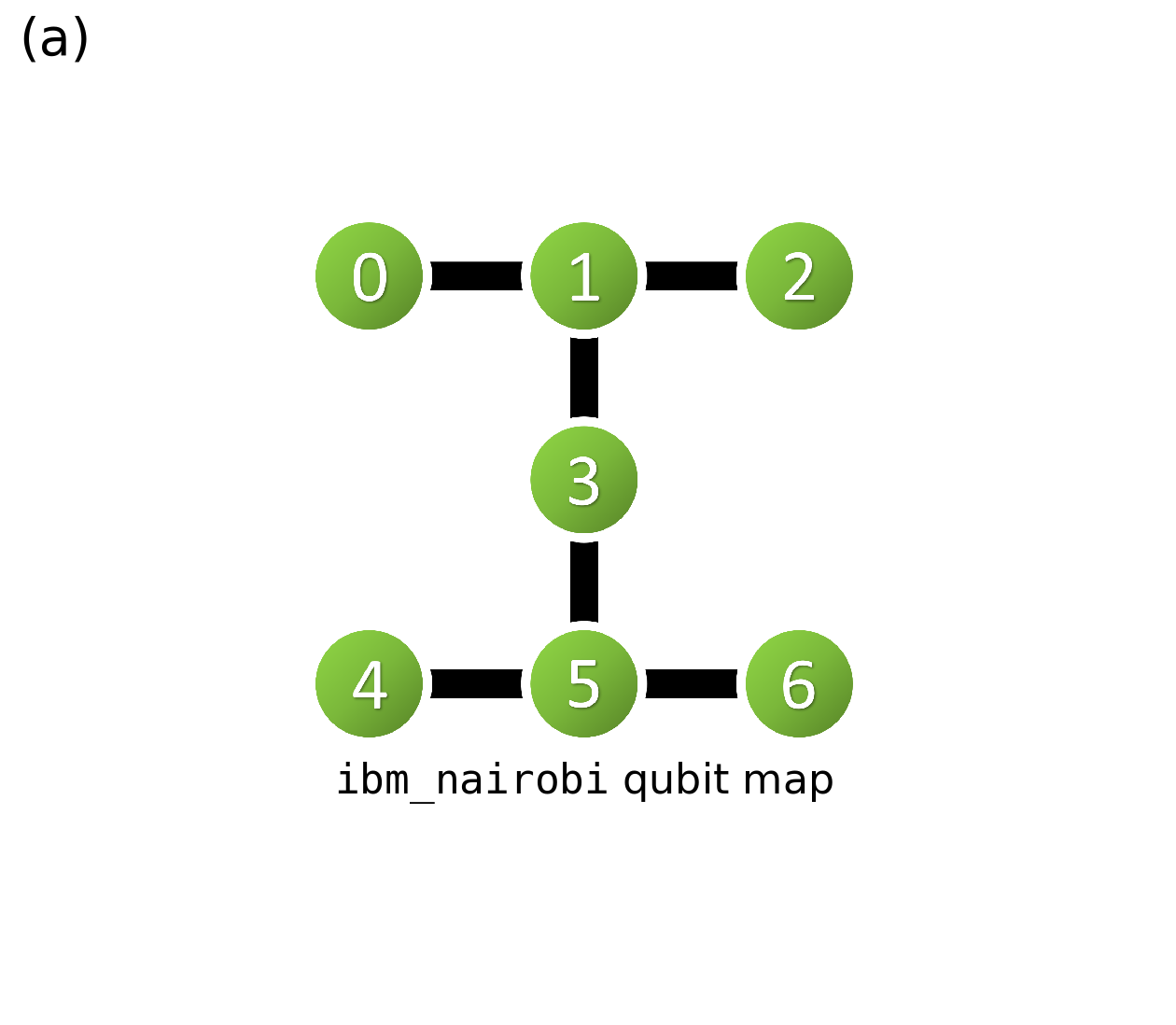}
\includegraphics[width=6.5cm]{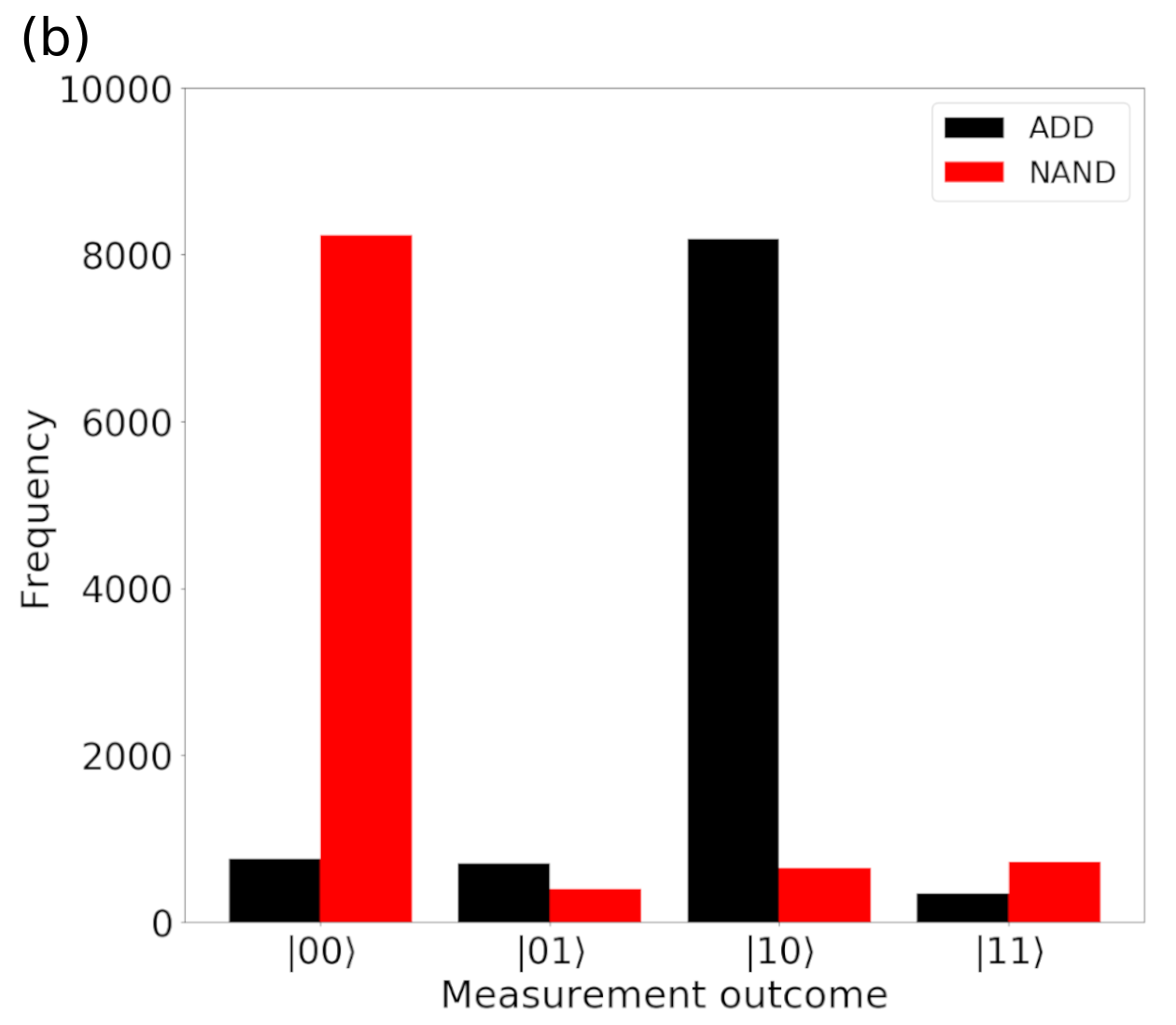}
\caption{\label{fig2} (a) The qubit map of 7-qubit \emph{ibm\_nairobi} quantum computer. (b) The measurement outcome (in order of $\mathrm{\ket{c_1 c_0}}$) versus frequency chart for arithmetic ADD (black) and logic NAND (red) operations on the QFT-based 2-input qALU.}
\end{figure}

In addition, we executed the qALU shown in Fig.~\ref{fig1} on a real quantum computer. We use 7-qubit \emph{ibm\_nairobi} quantum computer with the calibration data presented in appendix\,A. Further, the QFT-based qALU circuit is reconstructed in pre-process of IBM quantum computer to adapt the qubit map (topology) represented in Fig.~\ref{fig2}(a) and supported logic gates (CX, ID, RZ, SX and X) by \emph{ibm\_nairobi} quantum computer. In our demonstration, we set the initial numbers encoded as $\mathrm{\ket{q_0}=\ket{1}}$, $\mathrm{\ket{q_1}=\ket{1}}$ for two quantum channels and the operation-select ($\mathrm{\ket{S}}$) qubit of qALU is set as $\mathrm{\ket{0}}$ and $\mathrm{\ket{1}}$ for ADD and NAND operations, respecively. For the given numbers, the Fig.~\ref{fig2}(b) shows the results are $c_0=0$, $c_1=1$  and $c_0=0$, $c_1=0$ with the given probabilities $\mathrm{\approx \%82}$ for ADD (black) and NAND (red) operations, respectively.
\subsection{QFT based one-bit 4-input qALU}
\begin{figure}[!ht]\centering
\includegraphics[width=.95\linewidth]{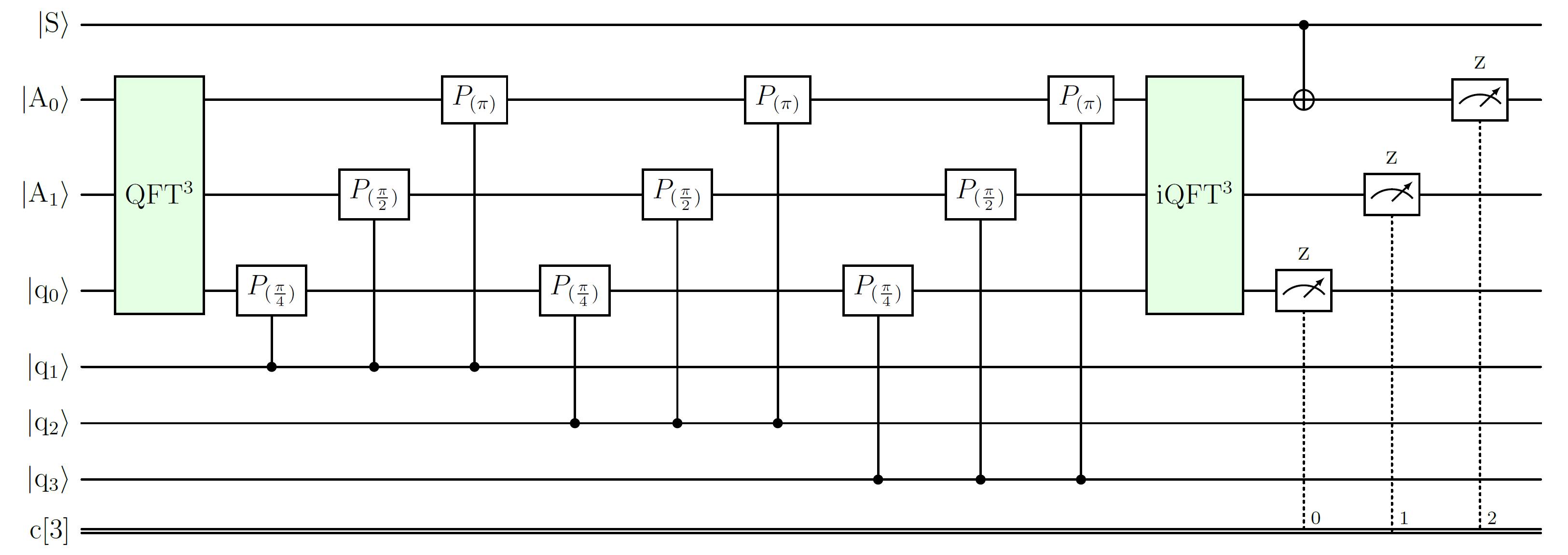}
\caption{\label{fig3} The quantum circuit of the QFT-based 4-input qALU. The $\mathrm{QFT^{3}}$ represents the 3 qubit QFT circuit (see Fig.~\ref{fig0}), and $\mathrm{iQFT^{3}}$ is the inverse QFT. The qALU performs parallel the arithmetic ADD and logic NAND operations on all input qubits.}
\end{figure}

The circuit shown in Fig.~\ref{fig1} only drives the two inputs (numbers), with each input holding one-bit (in encoded qubit) information. For the multi-qubit input, the algorithm can be executed $\mathrm{n-1}$ times sequentially within a loop by adding the previous sum to the new sum in order to add up $\mathrm{n}$-qubit numbers in two quantum channels. This generic approach is known as serial (sequential) addition. On the contrary, in QFT-based qALU, it is possible to apply the arithmetic ADD and logic NAND operations to multi-channel (input) qubits in Fourier space, simultaneously. The Fig.~\ref{fig3} represents the QFT-based one-bit 4-input qALU which drives the arithmetic ADD and logic NAND operations on four qubits (each qubit holds the one-bit numbers) without using any similar loop defined above. In this circuit, we have four encoded input qubits denoted $\mathrm{\ket{q_n}}$ where $n=\{0,1,2,3\}$, two ancillary qubits $\mathrm{\ket{A_0}}$, $\mathrm{\ket{A_1}}$ and operation-select $\mathrm{\ket{S}}$ qubit (see Fig.~\ref{fig3}).

\begin{figure}[!htb]\centering
\includegraphics[width=7.5cm]{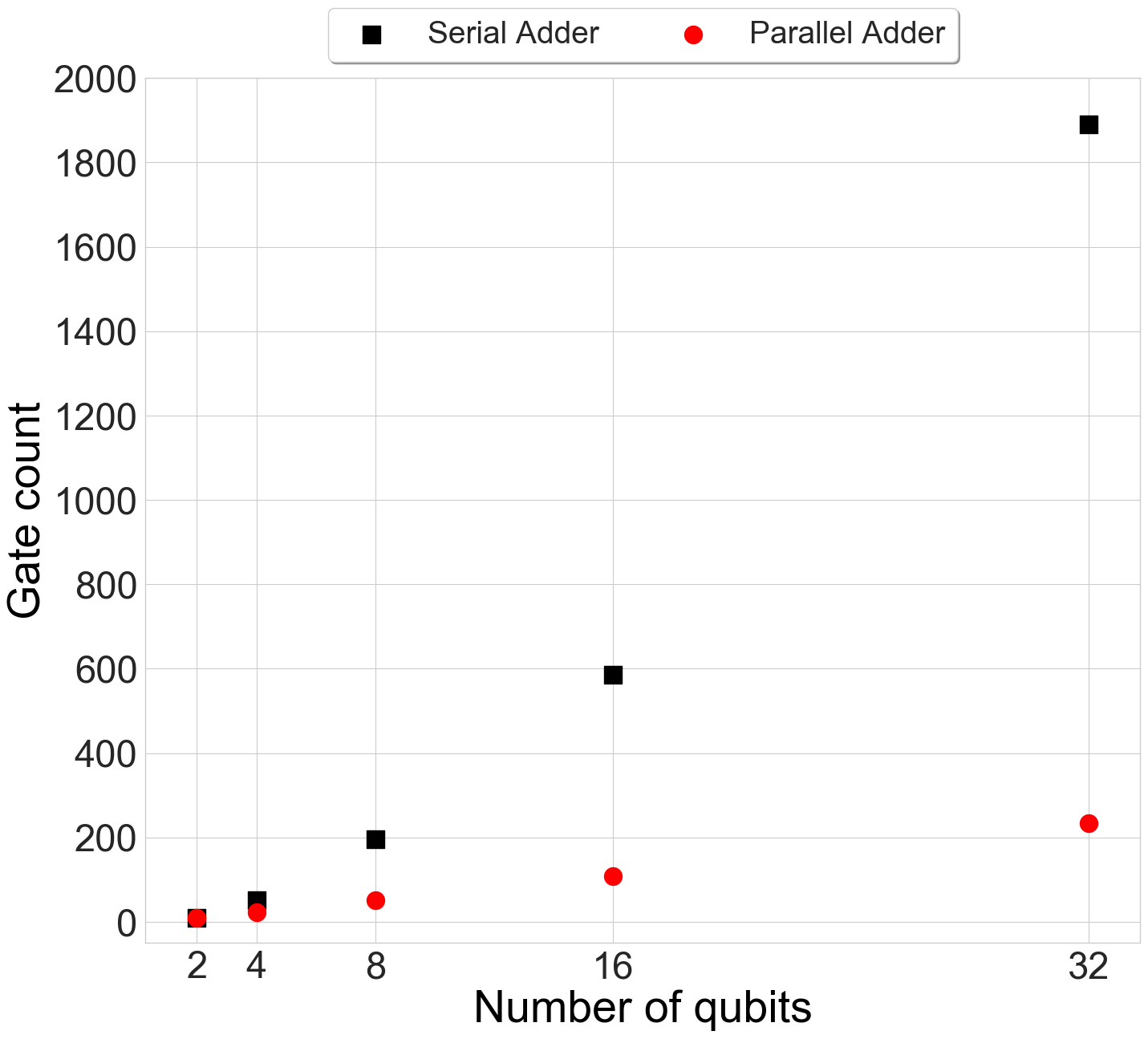}
\caption{\label{fig4} Number of the quantum logic gates required for QFT-based arithmetic ADD operation for serial (black) and parallel (red) architectures, respectively.}
\end{figure}

Lastly, we calculated the number of the quantum logic gates required for the arithmetic ADD opeation for serial (sequential) and parallel adder circuit designs. In Fig.~\ref{fig4}, the number of quantum logic gates required for the QFT-based arithmetic ADD operation circuit is calculated for $2^n$ one-bit numbers for serial (black) and parallel (red) designs are shown. It is evident from the figure that, as the number of input qubits increases, the parallel design constructed in this study shows a decrease in the number of logic gates compared to the serial addition. This is important for enhancing the noise effects on the quantum computing hardwares. Additionally, it can be predicted that the QFT-based multi-input qALU will perform operations faster compared to the two-input serial design.

\section{\label{conc}Conclusion and discussion}
In the presented research, we have developed a quantum circuit architecture for the QFT-based primitive qALU that performs an arithmetic ADD and logic NAND operations. We showed two circuits in Figs.~\ref{fig1} and~\ref{fig3} that realize the ALU operations on two-inputs and four-inputs, each input holds the number encoded on qubit, respectively. In particular, we implemented the qALU on a 7-qubit \emph{ibm\_nairobi} quantum computer utilizing the calibration values table provided in the appendix\,A. The QFT-based qALU circuit has a scalable architecture for handling large numbers encoded in multiple qubits.

The speed of matrix operations is very critical for graphic handling process. Also, it is very important for scientific research, medicine and gaming sectors. Unlike GPUs (Graphical Processing Units) which uses multi-ALU can be run in parallel, known as CUDA architecture~\cite{Glaskowsky09}, only a single QFT-based multi-input qALU can be performed the large matrix operations.

\begin{figure}[!ht]\centering
\includegraphics[width=6.5cm]{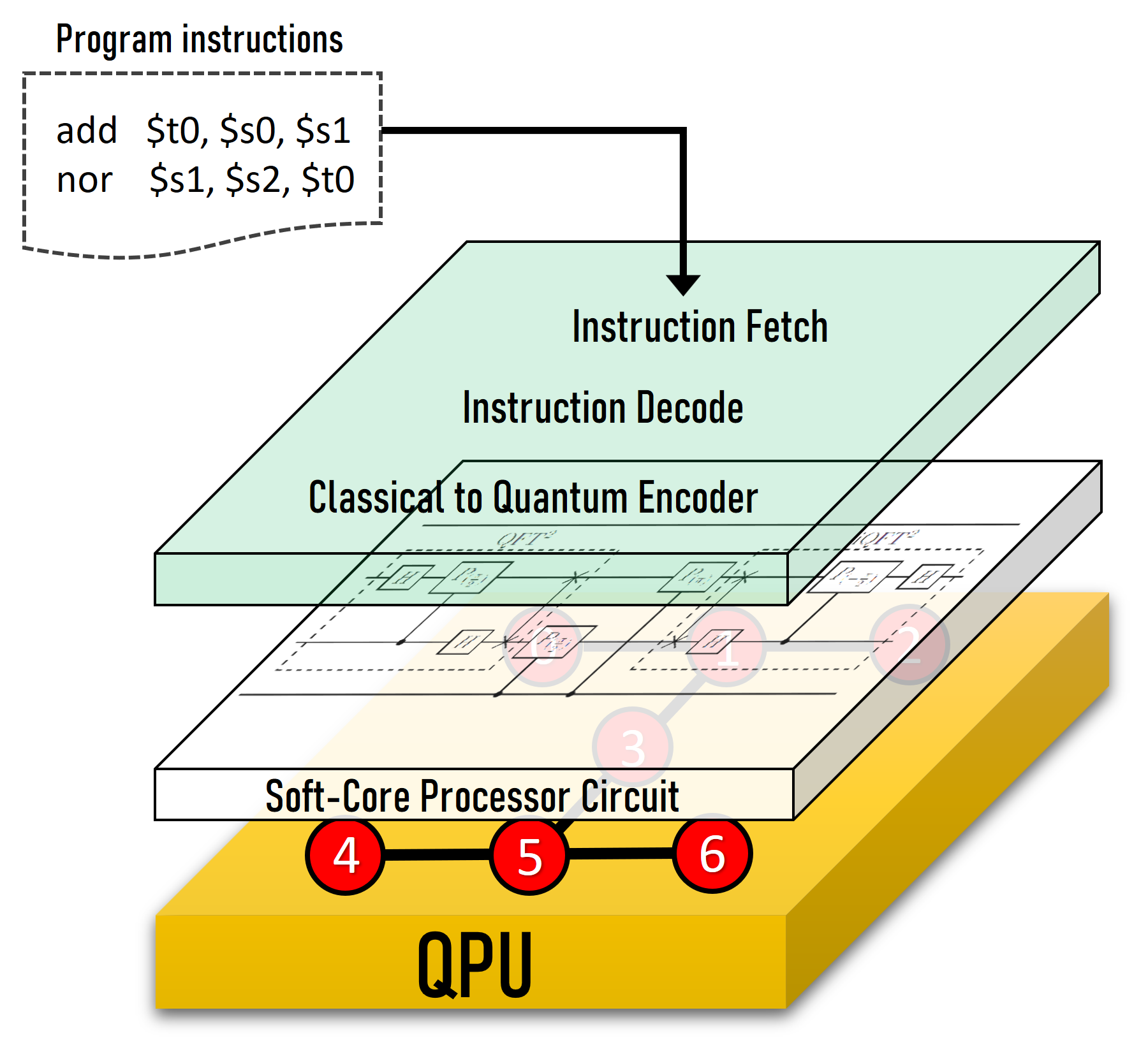}
\caption{\label{fig5} A possible architecture of the soft-core processor on quantum computer hardware (or field programmable spin arrays~\cite{Wang23}) is shown. The QFT-based qALU serves as the fundamental component of the soft-core processor. In the top layer, the program instructions are fetched and decoded, and then encoded on the qubits. In the middle layer, the soft-core processor algorithm is applied on the QPU (bottom layer) by utilizing quantum logic gates.}
\end{figure}

In addition, the arithmetic logic unit (ALU) is the common unit for executing program instructions in CPU architectures. In this research, it is shown that conventional instruction sets can be performed on quantum hardware, such as MIPS instruction for addition: \emph{add \$t0, \$s0, \$s1}. This makes it possible to construct software core (soft-core) processors that drive the tasks of conventional CPUs on quantum processing unit (QPU) hardwares as illustrated in Fig.~\ref{fig5}. It means that one can directly execute conventional computer programs on quantum computers without making any changes to the program source code, which is typically written in programming languages such as Python. The soft-core processor can also be designed for a chosen standard instruction set architecture, such as RISC (Reduced Instruction Set Computer), on a QPU. This is conceptually similar to a soft-core processor designed on a field-programmable gate array (FGPA)~\cite{Baklouti14,Amiri17,Nolting19,Finotti21,Titare23}, but there are significant technical differences between the classical and quantum based versions.

\ack
The authors acknowledge support from the Scientific and Technological Research Council of Turkey (T\"{U}B\.{I}TAK-Grant No. 122F298). We acknowledge the access to advanced services provided by the IBM Quantum Researchers Program.


\appendix
\section{Calibration data}
The calibration data of the 7-qubit \emph{ibm\_nairobi} quantum computer that we have used for the demonstration of the QFT-based one-bit 2-input qALU circuit, is presented in table~\ref{calib_data}.
\begin{table}[h]
\caption{\label{calib_data}The calibtation data for 7-qubit \emph{ibm\_nairobi} quantum computer (Date: 2023-05-26 19:40:01 UTC).}
\begin{indented}
\item[]\begin{tabular}{@{}lllll}
\br
$\mathrm{Qubit}$&$\mathrm{T_1(\mu s)}$&&$\mathrm{T_2(\mu s)}$&$\mathrm{Frequency(GHz)}$\\
\mr
0&120.25  &&31.66  &5.260\\
1&117.67  &&87.83  &5.170\\
2&88.44   &&144.51 &5.274\\
3&92.69   &&62.81  &5.027\\
4&54.06   &&58.16  &5.177\\
5&114.08  &&20.73  &5.293\\
6&121     &&114.03 &5.129\\
\br
\end{tabular}
\end{indented}
\end{table}

\end{document}